\documentstyle[11pt,newpasp,twoside]{article}
\def\edcomment#1{\iffalse\marginpar{\raggedright\sl#1\/}\else\relax\fi}
\reversemarginpar

\def\be{\begin{equation}}
\def\ee{\end{equation}}
\def\gtsima{$\; \buildrel > \over \sim \;$}
\def\gsim{\lower.5ex\hbox{\gtsima}}
\def\hmpc{\,h^{-1}{\rm Mpc}}
\def\hkpc{\,h^{-1}{\rm kpc}}
\def\solar{\ifmmode_{\mathord\odot}\;\else$_{\mathord\odot}\;$\fi}
\def\msun{{\rm M}_{\solar}}
\def\drm{{\rm d}}

\begin{document}
\title{Spin Profile of Galactic Halos and Disk Formation}
\author{A. Dekel$^1$, J.S. Bullock$^2$, C. Porciani$^1$, A.V. Kravtsov$^2$,
        T.S. Kolatt$^1$, A.A. Klypin$^3$ \& J.R. Primack$^4$}
\affil{$^1$ The Hebrew University of Jerusalem\\
       $^2$ Ohio State University\\
       $^3$ New Mexico State University\\
       $^4$ University of California, Santa Cruz}

\begin{abstract}
We summarize recent developments in the study of the origin of halo
spin profiles and preliminary implications on disk formation.
The specific angular-momentum distributions within halos in N-body 
simulations match a universal shape, $M(<\!j)\!\propto\!j/(j_0\!+\!j)$.
It is characterized by a power law over most of the mass,
and one shape parameter in addition to the spin parameter $\lambda$.
The angular momentum tends to be aligned throughout the halo 
and of cylindrical symmetry.
Even if angular momentum is conserved during baryonic infall, the
resultant disk density profile is predicted to deviate from exponential,
with a denser core and an extended tail.
A slightly corrected version of the scaling relation due to linear 
tidal-torque theory is used to explain the origin of a typical power-law 
profile in shells, $j(M) \propto M^s$ with $s\gsim 1$.
While linear theory crudely predicts the amplitudes of halo spins,
it is not a good predictor of their directions.
Independently, mergers of halos are found to produce a similar profile
due to $j$ transfer from the orbit to the product halo
via dynamical friction and tidal stripping.
The halo spin is correlated with having a recent major merger, 
though this correlation is weakened by mass loss. These two effects
are due to a correlation between the spins of neighboring halos and their 
orbit, leading to prograde mergers.
\end{abstract}

%%%%%%%%%%%%%%%%%%%%%%%%%%%%
\section{Introduction}

Disk galaxies that form in cosmological hydro simulations are 
surprisingly smaller and of lower spin than observed disks, posing
an ``angular-momentum problem" (e.g. Navarro \& Steinmetz 1997, 2000).  
A key ingredient in
the origin of galactic rotation, which can be studied in detail,
is the distribution of angular momentum in the dark halo within 
which baryonic infall is assumed to form the disk.
We describe here several advances in the investigation of the origin 
of halo angular momentum and its distribution, and preliminary 
implications on disk formation, based on high-resolution
$N$-body simulations and analytic approximations. The  
universal $j$ profile 
%(\se{prof}) 
(\S2) and the implied disk structure
%(\se{disk}) 
(\S3) are presented in Bullock et al.~(2000, BDK+).
The issues of tidal-torque theory 
%(\se{ttt}) 
(\S4) are studied by Porciani, Dekel, \& Hoffman (2000, PDH)
and the scaling relation is by Porciani \& Dekel (2000, PD).
The $j$ profile due to mergers 
%(\se{fric})
(\S5) is by Dekel \& Burkert (2000, DB).
The correlations with merger history 
%(\se{history})
(\S6) are to be described in Wechsler et al.~(2000, WBD+).

%%%%%%%%%%%%%%%%%%%%%%%%%%%%%%%%%%%%%%%
\section{A Universal Halo Spin Profile}
\label{sec:prof}

% ART simulations, halo finder
We use a high-resolution $N$-body simulation of the $\Lambda$CDM
cosmology ($\Omega_{\rm m}\!=\!0.3$) inside a $60\hmpc$ box,
with particles of $10^9\msun\!$ and force resolution of $2\hkpc$ (ART code,
Kravtsov et al.~1997).
Spherical halos are identified around density maxima using an NFW
density profile;
they contain a mass $M$ inside a virial radius $R$, defined such that
it encompasses a mean overdensity of 340. We focus here on 500 isolated halos
with $M>10^{12} h^{-1}\msun$.

%Profile
We find that the mass distribution of specific angular momentum $j$
is well fit by a {\it universal\,} profile:
\be
M(<j)= \mu M\, j /(j_0 +j) \ , \quad \mu > 1 \ .
\label{eq:prof}
%1
\ee
This is roughly a power law for $j < j_0$, which flattens off
if $j>j_0$.  The {\it shape\,} parameter $\mu$ correlates with 
the fraction of mass in the power-law regime, which always contains
more than half the mass. 
 
One can replace $j_0$ with a robust spin parameter 
$\lambda' \equiv J /(\sqrt{2} M V_{\rm c} R)$, where $V_{\rm c}^2 = G M/R$.
It coincides with the standard spin parameter $\lambda$
for an isothermal sphere and for an NFW profile with concentration $C\sim 10$.
The pairs $(\mu,j_0)$ and $(\mu,\lambda')$ are related via
\be
j_0\, b(\mu) = \sqrt{2} V_{\rm c} R\, \lambda' ,
\quad b(\mu) \equiv -\mu\, \ln(1-\mu^{-1}) -1 \ .
\label{eq:param}
%2
\ee
Either pair can fully characterize the $j$ profile.
%\fig{prof} 
Fig. 1 shows a few examples.
%
%\fig{mu-lambda}
Fig.~2 (left) shows a certain correlation between $\mu$ and $\lambda'$,
with a correlation coefficient $r=0.23$, meaning that
high-spin halos tend to have power-law profiles.

\begin{figure}
%1 a,b muspread.ps stack.ps
\vskip 5.5cm
{\includegraphics{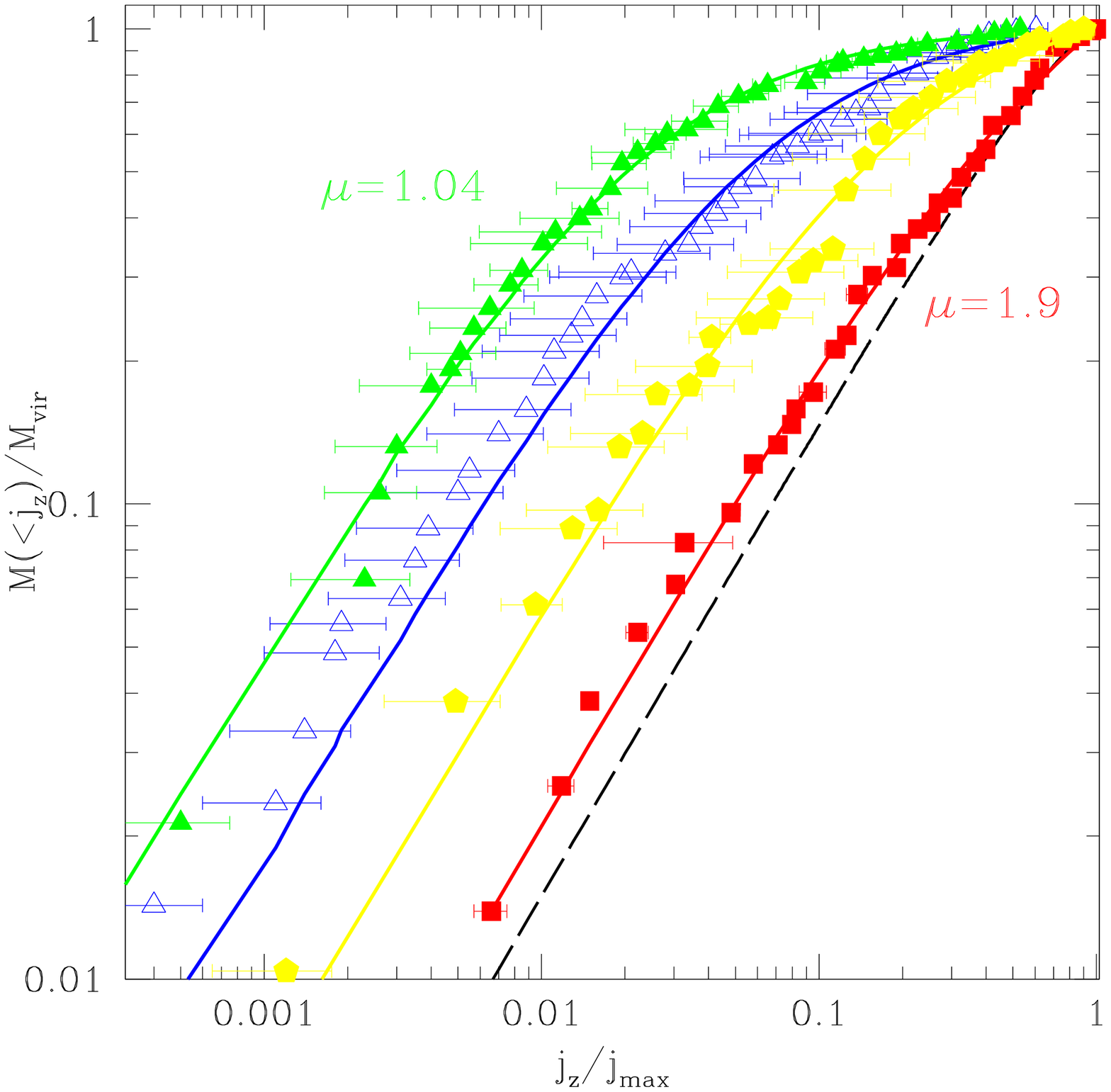}}
{\includegraphics{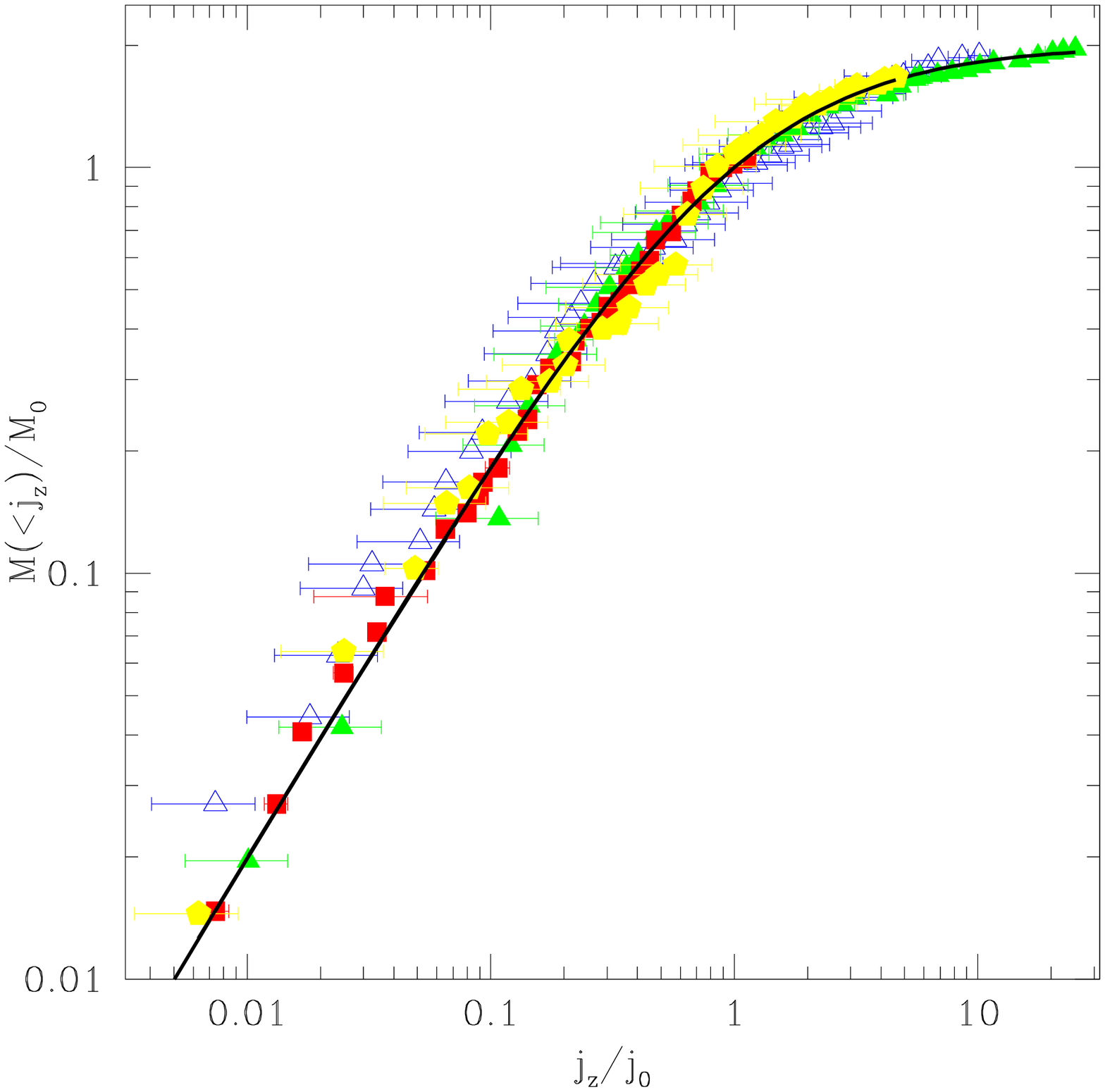}}
\caption{Mass distribution of specific angular momentum and the
corresponding fits by the universal profile for 4 representative
halos of different $\mu$ values.
{\bf Left:} scaled to coincide at the virial radius.
{\bf Right:} scaled to coincide at the bending point $(j_0,M_0)$. }
\label{fig:prof}
\end{figure}

%Alignment and cylindrical symmetry
We find that the direction of the angular-momentum vector tends to be aligned
throughout the halo; for more than half the mass, $j$ and $j_z$ typically 
differ by less than a factor of 2.  
The spatial symmetry of $j$ tends to be more cylindrical than spherical;
in 80\% of the halos, the mean value of $j$ (over longitude) 
decreases by less than a factor of 2 when moving  
away from the equatorial plane along lines parallel to the spin axis.

\begin{figure}
%2 mu.vs.lambda.ps log.log.sigma.ps
%Fig 2: mu-lambda correlated  
\vskip 5.5cm
{\includegraphics{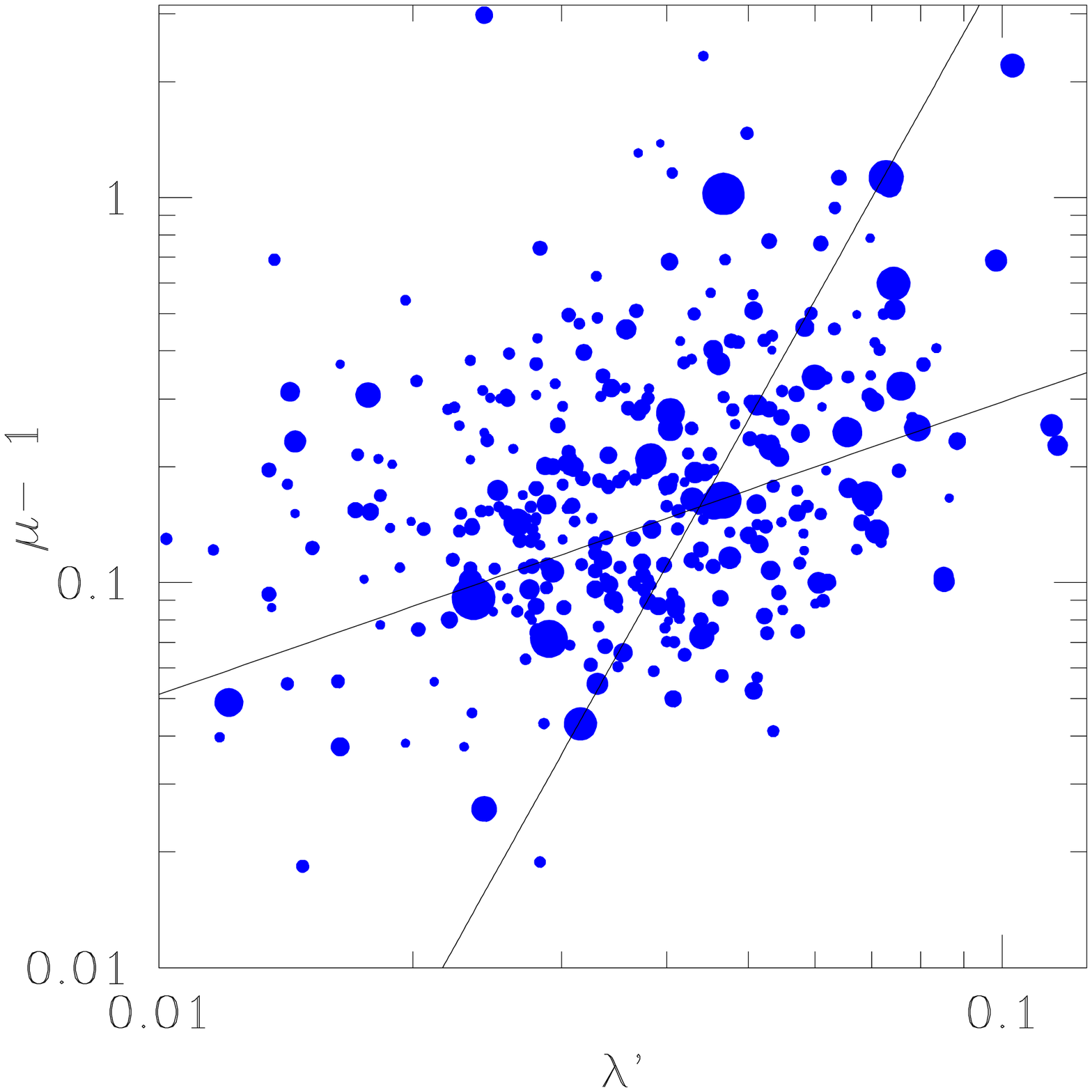}}
{\includegraphics{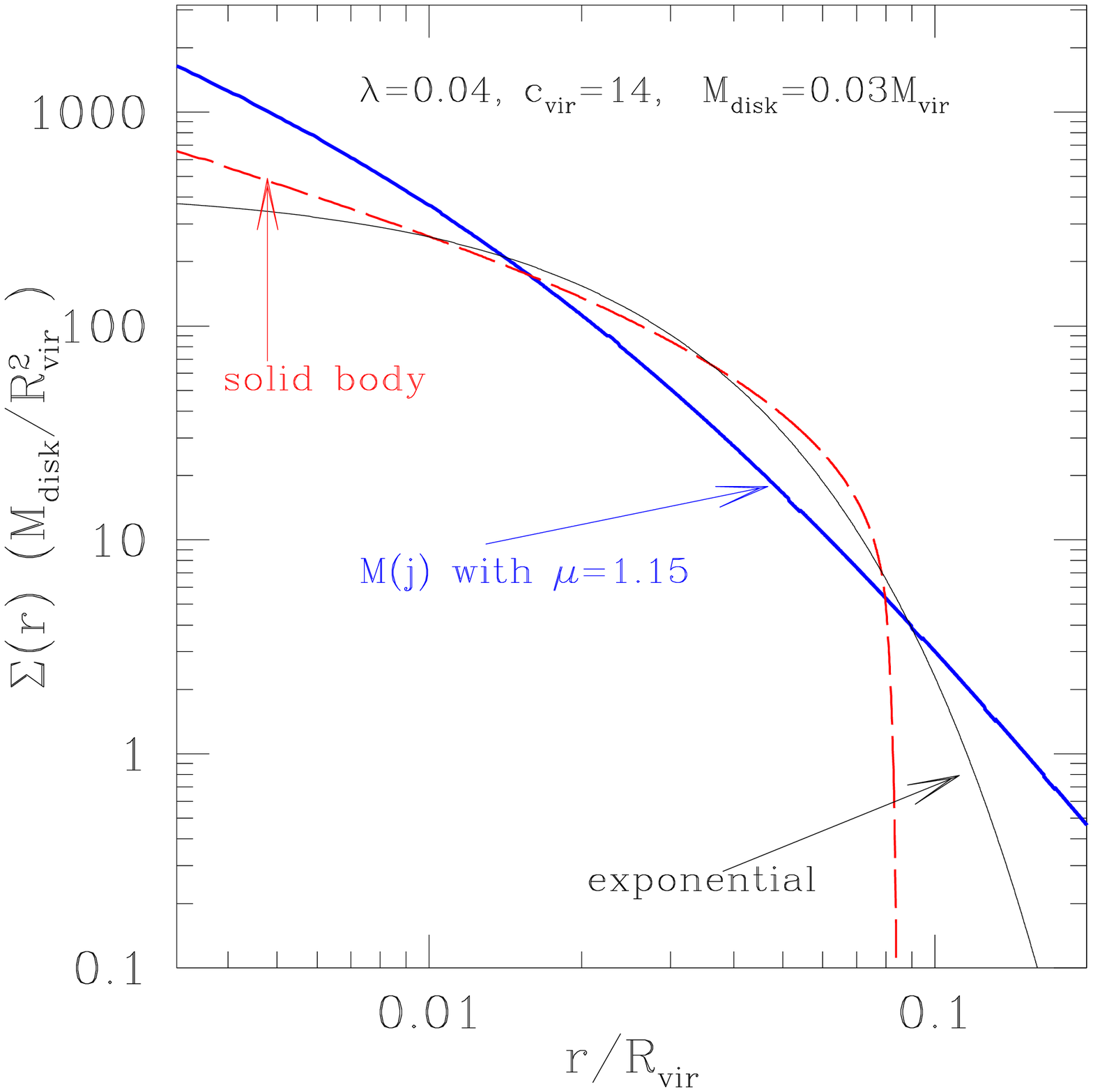}}
\caption{{\bf Left:} Correlation between angular-momentum parameters
$\mu$ and $\lambda'$. The symbol size
is inversely proportional to the error. Regression lines are shown; the
correlation coefficient is $r=0.23$.
{\bf Right:} Disk surface-density profile as implied by the halo $M(<\!j)$
with $j$ conservation.
Shown for comparison are an exponential disk and
the disk resulting from a uniform, solid-body rotating halo.
Note the density excess at small and large radii.}

\label{fig:mu-lambda}
\end{figure}

%%%%%%%%%%%%%%%%%%%%%%%%%%%%%%%%%%%%%%%
\section{Disk Structure: Core and Tail}
\label{sec:disk}

A simple model can convert the halo $M(<j)$  
into a disk density profile.
Assume that the gas initially traces the $j$ distribution 
of the halo, and that $j$ is conserved during 
the gas contraction into a rotation-supported disk of mass $fM$.
Then, using 
%\equ{prof}, 
Eq.~1, the disk mass inside radius $r$ is
$\propto j(r)/[j_0+j(r)]$. 
Under the assumption of an isothermal halo with a flat rotation curve,
one has $j(r)=rV_{\rm c}$, and then the disk surface-density profile is
\be
\Sigma_{\rm d} = {f\mu M \over 2\pi} { r_{\rm d} \over r (r_{\rm d} +r)^2} \ ,
\quad r_{\rm d} \equiv {\sqrt{2} \lambda' R \over b(\mu)} \ .
\label{eq:disk}
%3
\ee
The characteristic length scale $r_{\rm d}$ is based on 
%\equ{param}.
Eq.~2.
 
A more accurate numerical calculation using an NFW profile, 
that self-consistently solves
for the disk circular velocity at $r$ and takes into account the
adiabatic contraction of the halo in response to the baryonic infall,
leads to the typical profile shown in 
%\fig{disk}.
%\fig{mu_lambda}
Fig.~2 (right).
Compared to an exponential disk (which is similar to the profile
obtained from a uniform, solid-body rotating halo),
our resultant profile has a core and tail excess, with somewhat
higher densities both at small and large radii.
Since observed (stellar) disks are closer to exponential,
this result may either indicate that $j$ is not exactly preserved
during the collapse, or it may require some additional viscous processes
(e.g.~Lin \& Pringle 1987). 
In some cases, the obtained slow-rotating core may be associated with a bulge.
The extended disk, on the other hand, may be useful in explaining 
the damped Lyman-alpha systems observed at high redshift (Maller et al. 2000).

Our result for the limiting case of $j$ conservation clearly sharpens the
big ``angular-momentum problem", because $j$ transport from gas
to dark matter and from inside out can only worsen any agreement
with observed disks. It seems that one needs to appeal to some
process of gas removal from subclumps before they merge into the 
halo center, and thus reduce $j$ transport.

%%%%%%%%%%%%%%%%%%%%%%%%%%%%%%%%%%%%%%%%%%%%%%%%%%%
\section{Spin Profile based on Tidal Torque Theory}
\label{sec:ttt}

We find in the simulations that the halo $j$ profile in spherical shells 
is roughly a power law, $j(M) \propto M^s$, with $s$ distributed
like a Gaussian: $s=1.3 \pm 0.3$ (see an early indication by
Barnes \& Efstathiou 1987). Here $j$ is the specific angular momentum 
in a shell of radius $r$, and $M$ is the mass in the sphere encompassed 
by this shell.
This profile can be related to the distribution $M(<j)$ under the 
assumptions of 
spherical symmetry for the density and cylindrical symmetry for $j$.
Linear tidal-torque theory (TT) provides one hint for the origin of this
power law, as follows.

%Scaling --> j(M)
TT (e.g.~White 1984) implies that the $j$ gained by a halo 
at time $t$ before its turn-around is
\be
J_i(t) = a(t)^2 \dot D(t)\, \epsilon_{ijk}\, T_{jl}\, I_{lk} \ ,
\label{eq:ttt}
%4
\ee
where the time growth is from some fiducial initial time $t_{\rm i}$, 
$I_{lk}$ is the inertia tensor of the proto-halo at $t_{\rm i}$,
and $T_{jl}$ is the tidal tensor at the halo center, smoothed on 
the halo scale. This is based on assuming the Zel'dovich approximation
for the velocities inside the proto-halo, and a 2$^{\rm nd}$-order 
Taylor expansion of the potential (see PDH). We show elsewhere (PD)
that the standard scaling relation of TT 
should be slightly modified; it should read 
\be
j \propto t_{\rm c}\, \sigma(M)\, M^{2/3} \ ,
\label{eq:scaling}
%5
\ee
where $t_{\rm c}$ is the turn-around time of the halo 
(assuming that the relevant cosmology at $t_{\rm c}$ is still EdS)
and $\sigma(M)$ is the rms density fluctuation on scale $M$ at 
$t_{\rm i}$.

To obtain the spin profile we apply 
%\equ{scaling} 
Eq.~5 shell by shell.
The turn-around time of each shell is determined by 
$\bar\delta(M) D(t_{\rm c}) \sim 1$, where $\bar\delta(M)$ is the mean 
density inside $M$.
The densities within the different shells are correlated, such that,
for a Gaussian field, the typical density fluctuation profile
about a random point scales like 
$ \delta(r) \propto \xi(r)$,
where $\xi(r)$ is the linear two-point correlation function (Dekel 1981).
This is accurate to a few percent also around a high peak (Bardeen et
al.~1986, Fig.~8).
Thus, for a power-law power spectrum $P_k \propto k^n$ and a flat universe, 
we obtain for the $j$ profile within each halo
\be
j(M) \propto M^{2/3 + (3+n)/3} .
%6
\ee
This implies $1<s<4/3$ for $-2<n<-1$, the range appropriate for 
large galactic halos in a CDM spectrum, in pleasant agreement with
our finding for the simulated halos.
A similar analysis, using the accretion
rate into halos based on the Extended Press-Schechter approximation,
also explains a weak anti-correlation detected in the simulations
between $s$ and halo mass (BDK+).

%Validity of TT (amp, theta)
Using simulated halos (in GIF simulations), PDH find that TT
predicts an amplitude for the halo spin in the right ball-park.
This confirms the relevance of TT to the $j$ distribution in halos.
On the other hand, PDH find that the predicted spin direction is only
weakly correlated with the true direction, with an average deviation of
$53^\circ$.
This reflects large inaccuracies
in the smoothing of the tidal tensor and the truncation of the
Taylor expansion of the potential. This relative failure of TT in predicting
the spin direction should be born in mind when trying to use TT for
the analysis of spin-spin correlations and weak-lensing phenomena.

%%%%%%%%%%%%%%%%%%%%%%%%%%%%%%%%%%%%%
\section{Spin Profile due to Mergers}
\label{sec:fric}

\begin{figure}
%3 fric_c12.ps spinc_1.ps
\vskip 5.5cm
{\includegraphics{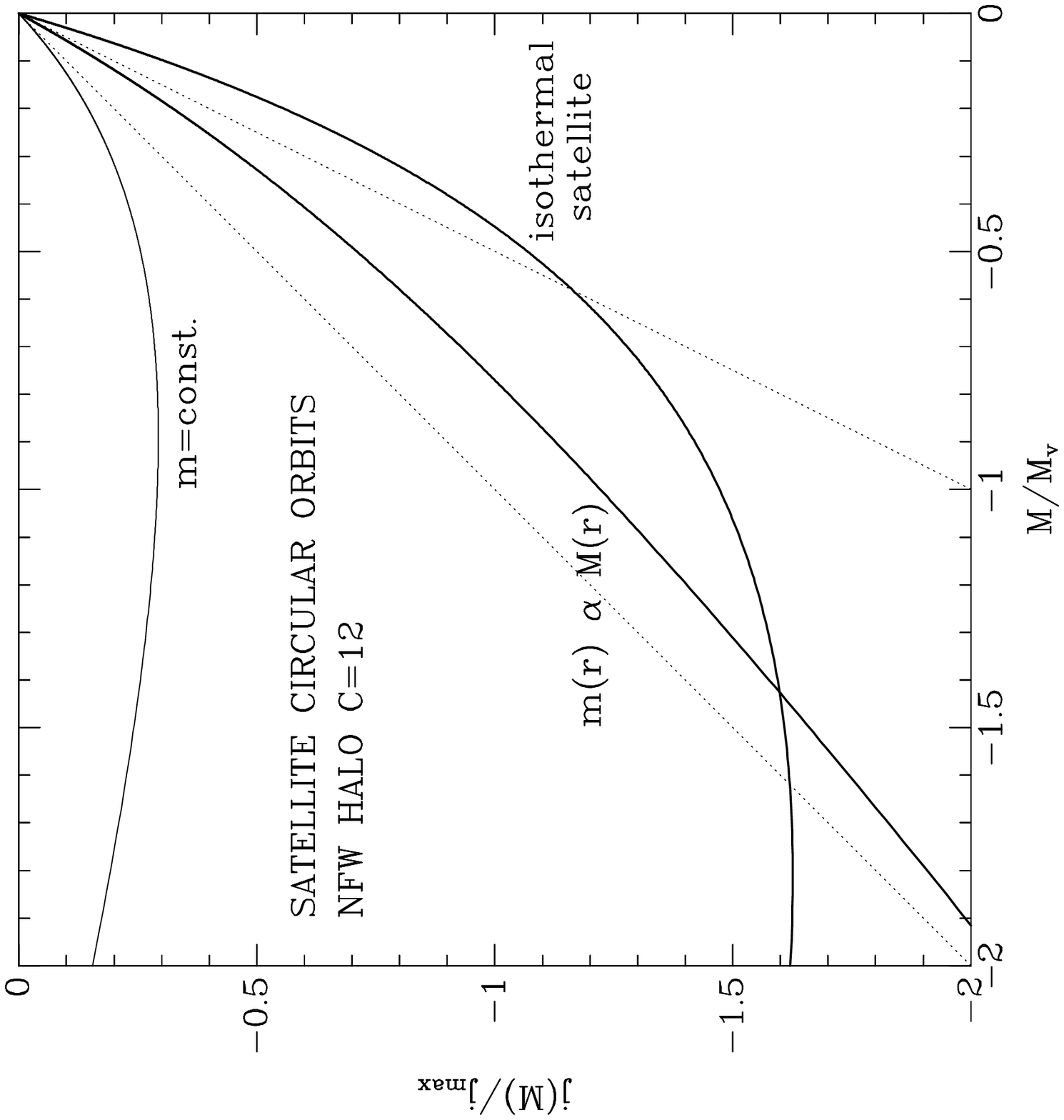}}
{\includegraphics{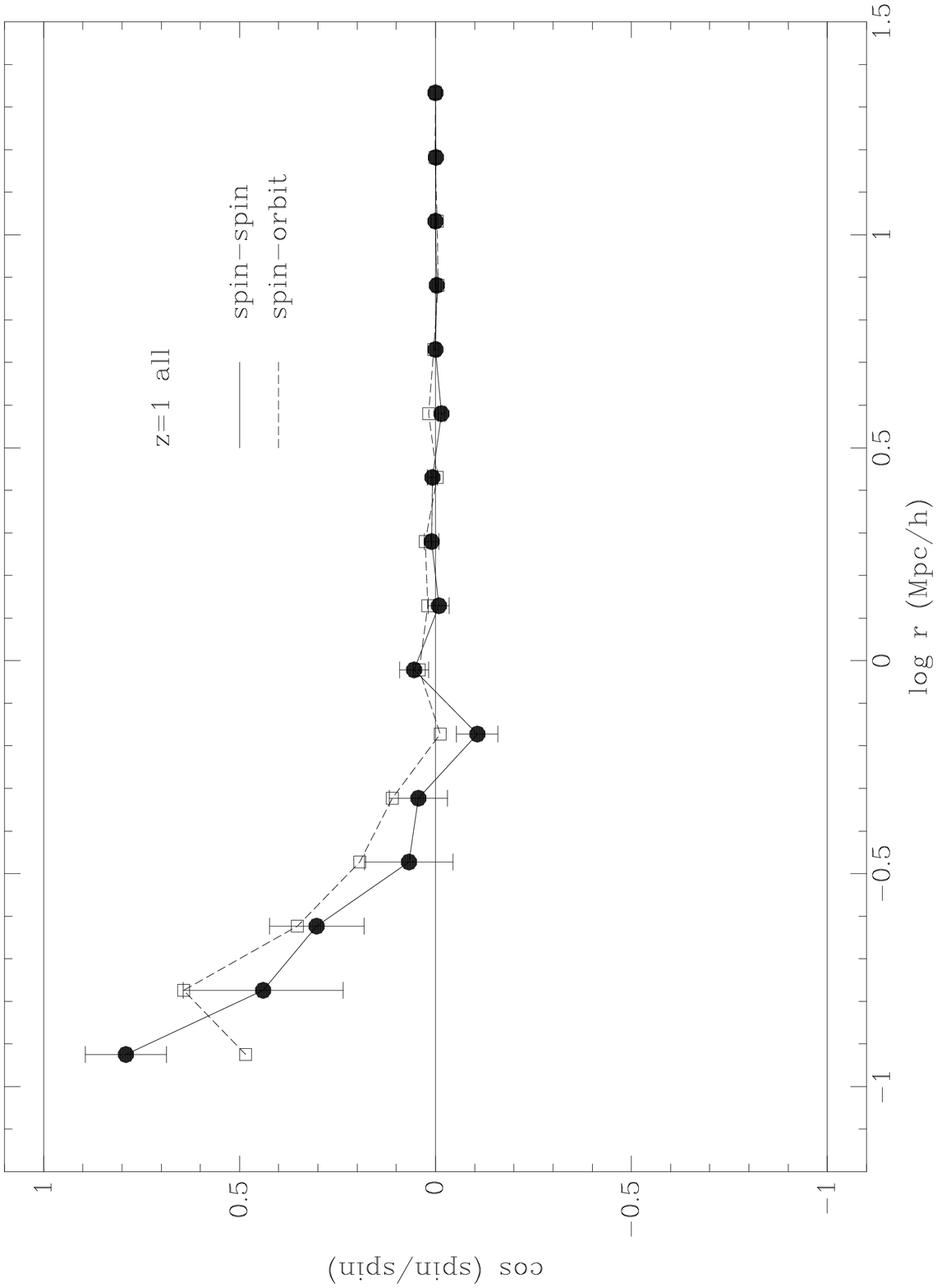}}
\caption{
{\bf Left:} Angular-momentum profile in spherical shells due to a minor
           merger of a satellite into an NFW halo, assuming circular
	   orbits and 3 different recipes for tidal mass stripping. 
	   The dotted lines are of slopes 1 and 2 for reference.
{\bf Right:} Halo spin correlations in a cosmological $N$-body simulation. 
Shown are the averages of cos(angle) between the spins and between spin 
and orbit for all pairs of halos separated by $r$.}
\label{fig:spinspin}
\end{figure}

A simple toy calculation for how $j$ is being deposited in the halo during
a merger provides another hint for the origin of the $j$ profile.
Consider a fixed halo and an incoming satellite of mass profiles 
$M(r)$ and $m(\ell)$ respectively.
Assume that as it moves in, the satellite is losing mass outside a 
tidal radius $\ell_{\rm t}$, determined at $r$ by
\be
{{m({\ell_{\rm t}})} / {\ell_{\rm t}^2}}={{\ell_{\rm t}\mu (r)} / {r^3}} \ ,
\quad    
\mu (r)=[ {2M(r)- r\, {\drm M /\drm r}} ] \ .
\label{eq:tidal}
%7
\ee
For isothermal halos this implies $m(r)\propto M(r) \propto r$.
Now assume that the satellite is spiraling in due to dynamical friction
in circular orbits,
and that the mass and $j$ are deposited locally. Then $j(r)$ is obtained
by averaging over shells,
\be
4\pi r^2 \rho (r)\, j(r) 
= m(r){\drm [rV_{\rm c}(r)] \over \drm r}
  +{\drm m(r) \over \drm r} rV_{\rm c}(r) \ .
\label{eq:Jdeposit}
%8
\ee
The $j$ transfer is due to both the slowdown of the satellite by 
dynamical friction and the direct tidal stripping of mass.
For an isothermal halo, we obtain
\be
j(M) \propto M \ .
%9
\ee

Fig. 3 (left) shows the $j$ profile for an NFW halo swallowing a
satellite that is loosing mass either according to the recipe
$m(r)\propto M(r)$ or based on Eq.~7 assuming that the satellite has 
an isothermal profile, compared to the case of no mass loss.
The general shape of $j(M)$ in the outer decade of mass, 
when realistic mass loss is considered, is reminiscent of the 
profiles detected in the cosmological simulation. 

Full $N$-body simulations of mergers, spanning a range of halo and 
collision parameters, confirm the robust production of such power-law 
like profiles (DB).

The resulting $j(M)$ from a sequence of minor mergers would be
a sum of similar contributions, each projected onto the direction of
the total net angular momentum (perhaps determined by the most 
major merger).

Since halo accretion histories typically reflect some combination of 
relatively quiescent mass accretion and more pronounced mergers,
the $j$ profiles of individual halos may reflect a combination of
the two processes explored above. We see that the profiles produced
by each of the two processes are, in general, of similar shape to the
profiles measured in the simulations.

%%%%%%%%%%%%%%%%%%%%%%%%%%%%%%%%%%%%%%
\section{Spin and Merger History}
\label{sec:history}

%lambda - history
Further investigations of halos in the ART and GIF simulations
(WBD+) reveal interesting correlations between halo spin and 
merger history. For example, halos that had a major merger in 
the last Gyr tend to have higher spins, $\langle \lambda \rangle = 0.055$,
in comparison with the average of 0.035 for all halos.
This is hardly surprising because most of the orbits of merging halos
are expected to be of large impact parameter and therefore to bring in
much angular momentum.

%spin loss
On the other hand, we find that major mergers also result in effective loss of
$j$, typically by a factor of $\sim 2$ of the
original orbital $j$.
This spin loss is associated with mass loss of the weakly bound
outer layers of the halos which are preferentially $j$-rich,
%\equ{prof}. 
Eq.~1. This spin loss somewhat weakens  the 
tendency for high spin in merger products.

%Spin-Spin
These effects may be correlated with the fact that major mergers 
tend to be prograde, with the spins of the two merging halos oriented
roughly parallel to each other and to the angular momentum of their 
mutual orbit. 
%\fig{spinspin} 
Fig.~3 (right) shows the average $cos(\theta)$ 
between the two spins and between the spin and orbit for pairs of 
halos at a given separation $r$. There is a significant alignment
at $r < 400\hkpc$, growing as a power law towards smaller distances.
Prograde encounters indeed explain both the high spin and the
high spin loss.  Being prograde they are more effective in transferring energy
from the orbit to internal motions and thus making the typical
impact parameter for mergers even larger. This, plus the 
coherency of progenitor
spins and orbital $j$, results in a high spin. But at the
same time a prograde encounter causes effective stripping of 
$j$-rich mass.

High spin due to major mergers seems not to go naturally
with the common wisdom that such mergers produce low-spin bulges and
elliptical galaxies.
However, it is the halo properties at $z\sim 1$ (say) 
rather than today that are likely to determine the $j$ content of the gas
which later falls in to form the disk.
Then, it is the halo merger history prior to $z\sim 1$ that is 
more relevant for the galactic spin, while later mergers govern 
the bulge-to-disk ratio. The high-spin halos at $z\sim 1$, those that 
were assembled and virialized not long before $z\sim 1$, may be the 
hosts of high-spin disks that are observed as low-surface-brightness 
galaxies (Dekel et al. 2001).

%%%%%%%%%%%%%%%%%%%%%%%%%%%%%%%%


\begin{references}
\reference
Bardeen, J.M., Bond, J.R., Kaiser, N., \& Szalay, A. S.  1986, ApJ, 304, 15
\reference
Barnes, J., \& Efstathiou, G. 1987, ApJ, 319, 575
\reference
Bullock, J.S., Dekel,A., Kolatt, T.S., Kravtsov, A.V., Klypin, A.A.,
   Porciani, C., \& Primack, J.R. 2000, ApJ, submitted, astro-ph/0011001 (BDK+)
\reference
Kravtsov, A., Klypin, A., \& Khokhlov, A.M. 1997, ApJS, 111, 73
\reference
Dekel, A. 1981, A\&A, 101, 79
\reference
Dekel, A., \& Burkert, A. 2000, in preparation (DB)
\reference
Dekel, A., et al. 2001, in preparation
\reference
Lin, D.N.C., \& Pringle, J.E 1987, ApJL, 320, 87
\reference
Maller, A.H., Prochaska, J.X., Somerville, R.S., \& Primack, J.R. 2000,
MNRAS, submitted (astro-ph/0002449)
\reference
Navarro, J.F., \& Steinmetz, M. 1997, ApJ, 478, 13
\reference
Navarro, J.F., \& Steinmetz, M. 2000, ApJ, 538, 477
\reference
Porciani, C., \& Dekel, A. 2000, submitted (PD)
\reference
Porciani, C., Dekel, A., \& Hoffman, Y. 2000, submitted (PDH)
\reference
Wechsler, R.H., Bullock, J.S., Dekel, A., Primack, J.R., Kravtsov, A.V.,
  \& Klypin, A.A. 2000, in preparation (WBD+)
\reference
White, S.D.M. 1984, MNRAS, 286, 38

\end{references}
\end{document}